\documentclass[reprint,
amsmath,amssymb,
aps,
prfluids,
url=false,
title=true]{revtex4-2}

\usepackage{placeins}
\usepackage{adjustbox}
\usepackage{graphicx}
\usepackage{dcolumn}
\usepackage{bm}
\usepackage{hyperref}
\usepackage[capitalize]{cleveref}

\usepackage{mathtools}
\usepackage{color}

\newcommand{\jump}[1]{[\![#1]\!]}
\newcommand{\four}[1]{{#1}_{\boldsymbol{q}}}
\newcommand{\avg}[1]{\left<#1\right>}

\let\originalleft\left
\let\originalright\right
\newcommand{\leftR}{\mathopen{}\mathclose\bgroup\originalleft}
\newcommand{\rightR}{\aftergroup\egroup\originalright}

\begin{document}

\preprint{APS/123-QED}

\title{Finite Membrane Thickness Influences Hydrodynamics on the Nanoscale}

\author{Zachary G. Lipel}
 \email{zl4808@princeton.edu}
 \affiliation{Department of Chemical and Biomolecular Engineering, University of California, Berkeley, California 94720, USA \\ Department of Chemical and Biological Engineering, Princeton University, Princeton, NJ 08544}
\author{Yannick A. D. Omar}%
 \email{yadomar@mit.edu}
\affiliation{Department of Chemical and Biomolecular Engineering, University of California, Berkeley, California 94720, USA \\ Department of Chemical Engineering, Massachusetts Institute of Technology, Massachusetts 02139, USA}%

\author{Dimitrios Fraggedakis}%
 \email{dimfraged@gmail.com \& dfrag@princeton.edu}
\affiliation{Department of Chemical and Biological Engineering, Princeton University, Princeton, NJ 08544}

\date{05/14/2025}

\begin{abstract}
Many lipid membrane-mediated transport processes--such as mechanically gated channel activation and solute transport--involve structural and dynamical features on membrane thickness length scales.
Most existing membrane models, however, tend to adopt (quasi-)two-dimensional descriptions that neglect thickness-dependent phenomena relevant to internal membrane mechanics, and thus do not fully account for the complex coupling of lipid membranes with their surrounding fluid media.
Therefore, explicitly incorporating membrane thickness effects in lipid membrane models enables a more accurate description of the influence of membrane/fluid coupling on transport phenomena in the vicinity of the bilayer surfaces.
Here, we present a continuum model for membrane fluctuations that accounts for finite membrane thickness and resolves hydrodynamic interactions between the bilayer and its surrounding fluid.
By applying linear response analysis, we observe that membrane thickness-mediated effects, such as bending-induced lipid reorientations, can generate shear flows close to the membrane surface that slow down the relaxation of nanometer scale shape fluctuations.
Additionally, we reveal the emergence of pressure inversion and flow reversal near the membrane interfaces, accompanied by localized stagnation points.
Among these, extensional stagnation points give rise to a novel mode of bulk dissipation, originating from bending-induced compression and expansion of the membrane surfaces and their coupling to shear stresses in the fluid.
Our findings identify membrane thickness as a key factor in nanoscale hydrodynamics and suggest that its effects may be detectable in fluctuation spectra and can be relevant to interfacial processes such as solute permeability and contact with solid boundaries or other membranes.

\end{abstract}

\maketitle

\textit{Introduction}---Lipid bilayers form the boundaries of cells and organelles, playing an important role in processes such as endocytosis, cellular adhesion, and action potential propagation~\cite{Iwasa1980,Watanabe2013,Shi2015,carlson_protein_2015,ling2020high}. 
Bilayers are commonly modeled as two-dimensional surfaces \cite{Helfrich1973,steigmann1999fluid,deserno2015fluid,sahu2017irreversible}, a description that has enabled the study of processes across a wide range of length scales—from membrane budding during endocytosis~\cite{liu2006endocytic,agrawal2009modeling,dmitrieff2015membrane,omar2020nonaxisymmetric} to the diffusion of individual proteins embedded in membranes~\cite{saffman1975brownian,agrawal2011model,Samanta2021}.
However, two-dimensional models cannot resolve the internal mechanics and hydrodynamic coupling that emerge at nanoscales, where the membrane thickness becomes the primary intrinsic length scale.
For example, thickness deformations are known to mediate protein function through hydrophobic mismatch~\cite{phillips2009emerging}, short-wavelength membrane fluctuations deviate from two-dimensional predictions due to finite thickness fluctuations~\cite{woodka2012lipid}, and membrane permeability to small solutes decreases systematically with increasing thickness~\cite{frallicciardi_membrane_2022}. 
To accurately describe the physics near the membrane surfaces at these scales, models should explicitly incorporate membrane thickness. \par

Building upon existing two-dimensional membrane descriptions, finite thickness effects are usually incorporated through phenomenological approaches that are inherently based on the Helfrich free energy~\cite{seifert1993viscous,evans1994hidden,merkel1989molecular,yeung1995unexpected,watson2011intermediate,fournier2015hydrodynamics,terzi2017novel,terzi_consistent_quad_2019,Levine_DeterminantsBending_2014,hamm2000elastic,pinigin2020additional,deseri2008derivation,Faizi2024}. 
Such approaches have been able to explore the relaxation dynamics of membrane-fluid systems~\cite{fournier2015hydrodynamics,rahimi2012shape} and reveal signatures of the membrane thickness on the nanoscopic density fluctuations, relevant to lipid bilayer structure factor measurements~\cite{watson2011intermediate,kelley2023nanoscale,nagao2017probing,Faizi2024}.
While these extensions of the Helfrich energy represent progress in modeling finite thickness effects on membrane dynamics, they do not fully account for membrane-fluid coupling consistent with three-dimensional continuum mechanics.\par

In this Letter, we present a finite thickness membrane model that follows from a self-consistent continuum mechanics framework and describes fluctuations of bilayers in contact with fluid reservoirs.
We find that shape fluctuations on the order of $1-10$ nm are affected by membrane thickness effects, exhibiting slower decay rates than predicted by two-dimensional theories.
At these length scales, we show that in-plane surface motion caused by lipid reorientations leads to pressure inversion in the bulk and the formation of two distinct types of stagnation points near the membrane: circulatory points, which do not contribute to dissipation, and extensional points, which give rise to viscous energy loss.
This latter class is characterized by bulk fluid dissipation driven by compression and expansion of the finite thickness membrane surfaces.

\begin{figure*}[t]
\centering
\includegraphics[width=1.0\linewidth]{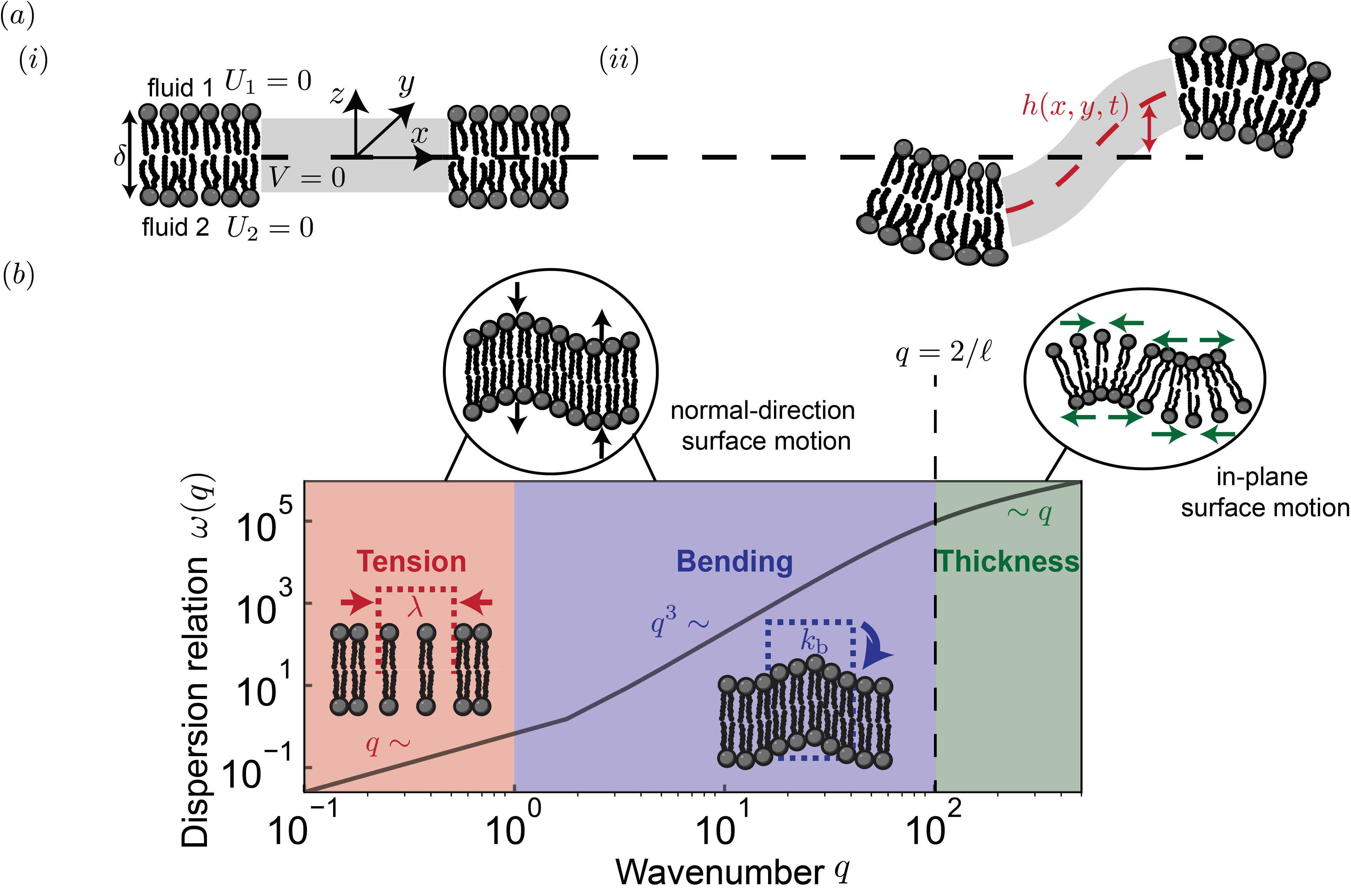}
\caption{(a)--(i) The base state of the linear response problem. 
The membrane is stationary and embedded in a quiescent fluid. 
(a)--(ii) The perturbed state, where the membrane mid-surface is displaced by the height field $h(x,y, t)$. 
(b) The dispersion relation for finite thickness, nearly flat membranes, as given by \cref{eq:dispersion_relation}. 
The tension regime crosses over into bending at $q_1 = \sqrt{2\Gamma} = 1.14$ and the viscous crossover occurs at $q_2 = 2/\ell = 100$, where where $\Gamma = \Lambda_0 L^2/k_\mathrm{b}$ and $\ell = \delta/L$. 
The elastic response is characterized by a tension that maintains mid-surface incompressibility and a bending modulus that penalizes moments generated about the membrane due to curvature. 
In the tension and bending regimes, motion at the membrane surfaces is primarily normal to the mid-surface and corresponds to height decay, whereas in the thickness regime, in-plane motion plays a dominant role in governing dissipation, arising from density relaxation.
Therefore, the viscous mechanism that dissipates membrane fluctuations transitions from normal to in-plane drag around the wavenumber $q=\ell/2$.
This shift in dissipation mechanism is only captured by a finite thickness theory that explicitly evaluates boundary conditions at the membrane surfaces, where the strongest bulk flows at the membrane interface occur in-plane at high wavenumbers. 
We choose as physical parameters $L = 200~\mathrm{nm}$, $k_\mathrm{b}=62~\mathrm{pN\cdot nm}$, $\Lambda_0=10^{-3}~\mathrm{pN\cdot nm^{-1}}$, and $\delta = 4~\mathrm{nm}$, and $\mu^\mathrm{b}=10^{-3}~\mathrm{pN\cdot nm^{-2}\cdot \mu s}$ \cite{Alberts_Bio, Phillips2018}.}
\label{fig:intro_fig}
\end{figure*}

\textit{Theory}—We consider a lipid bilayer of finite thickness $\delta$ immersed in a quiescent, incompressible viscous fluid. The bilayer is initially flat with its mid-surface located at $z=0$, as shown in \cref{fig:intro_fig}~(a)--(i). Physically, the system experiences thermal fluctuations that perturb the membrane shape and drive flows both within the membrane and in the surrounding bulk fluids. The primary quantities of interest are the membrane deformation, the in-plane lipid motion, and the hydrodynamic stresses transmitted across the membrane surfaces.

Our objective is to model how finite membrane thickness affects the relaxation dynamics of membrane shape fluctuations.
In particular, we are interested in the role of differential shear stresses between the top and bottom membrane surfaces on the coupling to the surrounding fluid--phenomena not fully resolved in strictly two-dimensional membrane models.
To describe this situation, we use a continuum framework for lipid membranes—originally developed in Refs.~\cite{omar2023electrostatics,omar2023balance,omar2025constitutive}—which systematically incorporates thickness effects through a spectral expansion in the thickness direction, leading to the $(2+\delta)$-dimensional equations for lipid bilayers.
As shown in~\cref{fig:intro_fig}~(a)--(ii), we represent the membrane shape using the Monge gauge, $F(x,y,z,t) = h(x,y,t)-z$~\cite{Monge1807}, valid for small deviations from flatness.
Under these assumptions, the membrane in- and out-of-plane momentum balances (\cref{eq:inplane_incompr_eqn_no_spont_curv,eq:shape_eqn_incompressible}) and continuity equation (\cref{eq:cont_eqn_incompressible}) take the form (see SM Sec.~5.2)
\begin{subequations}
    \begin{align}
    \begin{split}\label{eq:inplane_incompr_eqn_no_spont_curv}
        0 =&~ \partial_\alpha\lambda
        +  \mu^\mathrm{m}\nabla^2_\mathrm{s} v^\alpha +\mu^\mathrm{b}\jump{\partial_z u^\alpha + \partial_\alpha u^z}~,
    \end{split}\\
    \begin{split}\label{eq:shape_eqn_incompressible}
        0 =&~ \Lambda_0 \nabla_\mathrm{s}^2 h  - \frac{1}{2} k_\mathrm{b}\nabla^4_s h 
        + \mu^{\mathrm{b}}\delta\partial_\alpha \Big(\avg{\partial_z u^\alpha + \partial_\alpha u^z}\Big) \\
        &~-\jump{p} + 2\mu^\mathrm{b}\jump{\partial_z u^z}~,
    \end{split}\\
    \label{eq:cont_eqn_incompressible}
    0 =&~\partial_x v_x + \partial_y v_y~,
    \end{align}   
\end{subequations}
where $\alpha \in \{x,y\}$, $\boldsymbol{v}$ and $\boldsymbol{u}$ are the membrane and bulk velocities, respectively, $\lambda$ is the in-plane membrane tension enforcing mid-surface incompressibility, $k_\mathrm{b}$ is the bending rigidity, $\mu^\mathrm{m}$ and $\mu^\mathrm{b}$ are the membrane and bulk viscosities, $\Lambda_0$ is the base-state membrane tension, and $\nabla_\mathrm{s} \equiv \partial_x \boldsymbol{e}_x + \partial_y \boldsymbol{e}_y$ is the surface gradient operator. 
The double bracket notation, $\jump{\cdot} \coloneqq (\cdot)^+ - (\cdot)^-$, represents the jump in quantities evaluated at the membrane surfaces, e.g. at $z=\pm \delta/2$, while the angled bracket notation, $2\avg{\cdot} \coloneqq (\cdot)^+ + (\cdot)^-$, indicates their average.
 The appearance of the average terms is a result of the membrane resistance to applied torque.
The finite thickness formulation allows for both nonzero shear stresses and velocity differences between the top and bottom membrane surface, generating modes of coupling to the bulk fluid that are inaccessible to strict two-dimensional models~\cite{steigmann1999fluid,deserno2015fluid,Helfrich1973,sahu2017irreversible,canham1970minimum}.
Lastly, the surrounding fluid obeys the incompressible Stokes equations~\cite{happel1983mechanics,kim_microhydrodynamics_1991}, appropriate for microbiological processes where the Reynolds number is often $\mathcal{O}(10^{-6})$~\cite{Purcell1977,watson2011intermediate}.

To nondimensionalize the system, we introduce characteristic scales: $L$ for length, $U$ for velocity, and $\tau = L/U$ for time. The dimensionless parameters that arise are the Föppl–von Kármán number $\Gamma \coloneqq \Lambda_0 L^2/k_\mathrm{b}$, comparing membrane tension to bending forces; the capillary number $\mathrm{Ca} \coloneqq \mu^\mathrm{b} U/\Lambda_0$, contrasting viscous forces with membrane tension; the Scriven-Love number $\mathrm{SL}\coloneqq \mu^{\mathrm{m}} U L/k_{\mathrm{b}}$ \cite{sahu2020geometry}, relating membrane viscous to bending forces; and the dimensionless membrane thickness $\ell = \delta/L$.
In dimensionless form, the governing equations become
\begin{subequations}
    \begin{align}
    \begin{split}\label{eq:inplane_incompr_eqn_no_spont_curv_nondim}
        0 =&~ \partial_\alpha\lambda
        +  \frac{\mathrm{SL}}{\Gamma}\nabla^2_s v^\alpha+\mathrm{Ca}\Big(\jump{\partial_z u^\alpha + \partial_\alpha u^z}\Big)~,
    \end{split}\\
    \begin{split}\label{eq:shape_eqn_incompressible_nondim}
        0 =&~ \nabla_s^2 h  - \frac{1}{2 \Gamma}\nabla^4_s h 
        + \mathrm{Ca}\Big(\ell\partial_\alpha \big<\partial_z u^\alpha + \partial_\alpha u^z\big> \\
        &~-\jump{p} + 2\jump{\partial_z u^z}\Big)~,
    \end{split}\\
    \label{eq:cont_eqn_incompressible_nondim}
    0 =&~\partial_x v_x + \partial_y v_y~,
    \end{align}
\end{subequations}

Relevant values for the material parameters found in \cref{eq:inplane_incompr_eqn_no_spont_curv,eq:shape_eqn_incompressible,eq:cont_eqn_incompressible} are $k_\mathrm{b}\sim \mathcal{O}(10^2)~\mathrm{pN\cdot nm}$~\cite{Pecreaux2004,Phillips2018,takatori2020active}, $\Lambda_0 \sim \mathcal{O}(10^{-4}-10^{-1})~\mathrm{pN\cdot nm^{-1}}$~\cite{Pecreaux2004,Dai1998,sahu2020geometry}, with length scales ranging from $\mathcal{O}(10^2)~\mathrm{nm}$ for vesicles to $\mathcal{O}(1)~\mathrm{\mu m}$ for cells~\cite{Purcell1977,sahu2020geometry,Alberts_Bio,Phillips2018}, and velocities from $\mathcal{O}(10^{-3})~\mathrm{nm/\mu s}$ for tube pulling~\cite{shi2018cell,leduc_cooperative_2004,sahu2020geometry} to $\mathcal{O}(10)~\mathrm{nm/\mu s}$ for bacterial gliding~\cite{nan_uncovering_2011,tchoufag_mechanisms_2019}. Taking the viscosity of water for the bulk fluids, the resulting dimensionless parameters typically satisfy $\Gamma\sim 10^{-2}-10^3$, $\mathrm{Ca}\sim 10^{-7}-10^2$, and $\ell\sim 10^{-3}-10^{-2}$.

To explore the membrane dynamics, we analyze the linear response of the system to small shape fluctuations. 
We expand all dynamical quantities into normal modes: $f(x,y,z,t) = \sum_{\boldsymbol{q}} f_{\boldsymbol{q}}(z)e^{i\boldsymbol{q}\cdot\boldsymbol{x} + \omega(q)t}$, where $\boldsymbol{q} = (q_x, q_y)$ is the dimensionless in-plane wavevector, $\omega(q)$ is the dispersion relation encoding the membrane dynamical response, and surface quantities such as the displacement field $h(x,y,t)$ are evaluated at $z=0$.
By applying classical no-slip boundary conditions between the membrane and the bulk fluids, while allowing for differential shear at the two membrane surfaces, and assuming decay of bulk velocities and pressures far from the membrane, we solve~\cref{eq:inplane_incompr_eqn_no_spont_curv_nondim,eq:shape_eqn_incompressible_nondim,eq:cont_eqn_incompressible_nondim} to find an expression for the dispersion relation $\omega(q)$ (see SM Sec.~5.2 for details),
\begin{equation}
    \omega(q) = -\frac{\frac{1}{2}q^3 + \Gamma q}{\Gamma\mathrm{Ca}\left(4 + q^2\ell^2\right)}~.
    \label{eq:dispersion_relation}
\end{equation}
The form of \cref{eq:dispersion_relation} explicitly includes membrane thickness effects in the denominator, i.e., $q^2\ell^2$, on the relaxation dynamics of the membrane shape, a result that we proceed to explore in greater detail in the following section of this letter.

\begin{figure*}[t]
\centering
\includegraphics[width=1.0\linewidth]{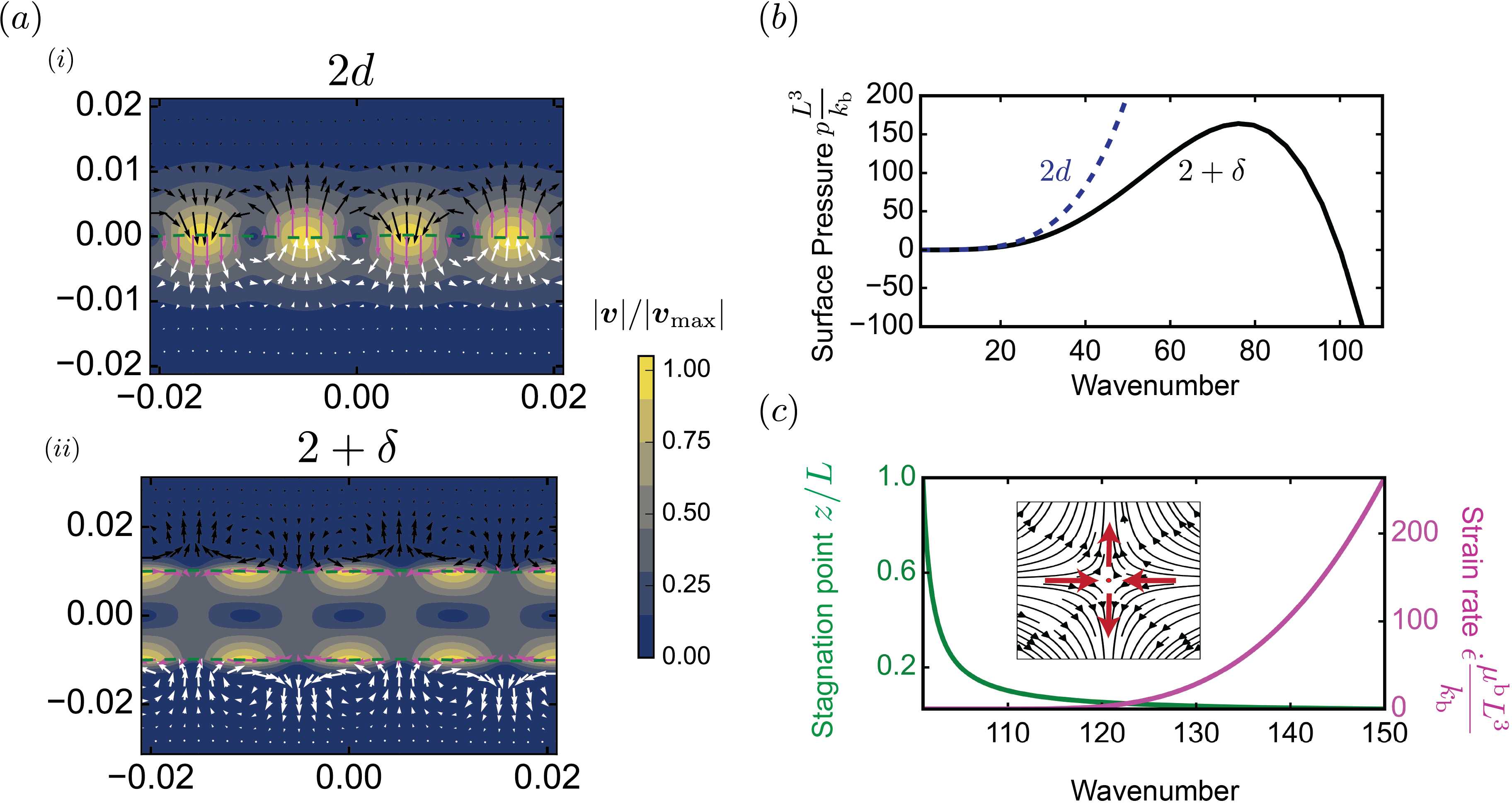}
\caption{\label{fig:results_fig}(a)--(i), (ii) The instantaneous ($t=0$) flow fields induced by a mode with $q= 300 \gg q_2$ for the two- and $(2+\delta)$-dimensional theories, respectively. For clarity and the sake of comparison, we do not show the intramembrane profiles in (ii). The dominant surface motion in the two-dimensional theory is normal to the membrane and corresponds to height field relaxation while in the $(2+\delta)$-dimensional theory it is tangential and corresponds to lipid reorientations. Additionally, vortices develop in the bulk when thickness is included. (b) The surface perturbed pressure at $(x,z) = (\pi/2q, \ell/2)$ as a function of wavenumber for the two- and $(2+\delta)$-dimensional theories. Inclusion of finite thickness modulates the pressure, which exhibits inversion at $q = q_2=2/\ell$. This signifies local flow direction reversal, which is attributed to bulk incompressibility and local in-plane flows in the thickness-dominated regime. (c) The extensional flow stagnation point position and strain rate as a function of wavenumber. We plot only for $q>q_2$, as the predicted stagnation point location is unphysical for lower wavenumbers. As the membrane surface in-plane deformations relax faster at higher wavenumbers, the stagnation points approach the membrane and their strain rate increases. The bulk circulations become increasingly localized near the membrane surface as the wavenumber increases. Including finite thickness allows us to resolve the hydrodynamics local to the membrane. The same parameters as in Fig.~\ref{fig:intro_fig} are used here. The capillary number $\mathrm{Ca}$ is set to $1$.}
\end{figure*}

\textit{Results}---
To understand the implications of membrane thickness on its dynamic response, we proceed to analyze the regimes of the dispersion relation $\omega(q)$, as shown in \cref{fig:intro_fig}~(b).
We observe three distinct regions in the dispersion relation, separated by the crossover wavenumbers $q_1 = \sqrt{2\Gamma} = 1.14$ and $q_2 = 2/\ell = 100$ for the choice of parameters in \cref{fig:intro_fig}. 
For $q \ll q_1$, membrane shape fluctuations are dominated by tension and relax according to $\omega(q) \sim q$.  
For $q_1 \ll q \ll q_2$, bending-dominated modes govern the relaxation dynamics, decaying as $\omega(q) \sim q^3$.
In the short wavelength regime where $q \gg q_2$, the mechanism for membrane relaxation shifts from viscous dissipation primarily via normal forces--such as pressure differences across the membrane--to dissipation dominated by in-plane forces arising from membrane surface shear stresses at the interfaces with the bulk, as illustrated in \cref{fig:intro_fig}--(b).
These shear stresses appear as $\mathcal{O}(\ell)$ terms in~\cref{eq:shape_eqn_incompressible} and develop due to lipid reorientation and resultant bulk tangential flows at the membrane-fluid interfaces. 
This in-plane viscous dissipation is explicitly captured by finite thickness theories but not by strict two-dimensional theories \cite{helfrich_undulations_1984, mutz_bending_1990, brochard_erythrocytes_1975, takatori2020active}, indicating a novel mechanism by which membrane thickness slows down the response to thermally induced fluctuations.

To understand these thickness effects on the hydrodynamic response for $q > q_2$, we consider $q = 300$ and examine the induced bulk flow fields near the membrane surfaces.
Figures~\ref{fig:results_fig}~(a)--(i) \& (ii) show the velocity vector fields along with their normalized magnitude for $\ell = 0.0$ (i.e. a two-dimensional membrane) and $\ell = 0.02$, respectively.
In \cref{fig:results_fig}~(a)--(i), we observe that the strictly two-dimensional membrane moves only in the normal direction. 
Minima in the initial perturbation push and drag bulk fluid in the positive $z$-direction, while maxima do so in the opposite direction, generating circular flow patterns around the membrane as a consequence of incompressibility in the bulk.
While we only show flow fields for $q=300$, the qualitative nature of the hydrodynamic response of the strict two-dimensional membrane is independent of the wavenumber (see SM Sec.~6.2).

Finite thickness is expected to affect the hydrodynamic response for $q > q_2$, as alluded to in the analysis of the dispersion relation.
This is evident in \cref{fig:results_fig}~(a)--(ii), where we observe the emergence of in-plane motion at the membrane-fluid interfaces.
This phenomenon arises due to membrane surface compression and expansion, which necessarily accompany bending. 
While curvature effects are included in both the strict two-dimensional and the $(2+\delta)$-dimensional theories, the latter improves upon the former by allowing us to resolve the three-dimensional nature of bending by incorporating finite thickness explicitly.
To explore the consequences of these thickness effects, we seek to understand the hydrodynamic response of the bulk fluid.

In the thickness-dominated regime, in-plane motion of the membrane surfaces induces in-plane fluid flow through no-slip conditions. 
Consequently, mass conservation in the bulk fluids necessitates the formation of circulations that appear near the top and bottom surfaces of finite thickness membranes. 
To further understand how membrane thickness affects the hydrodynamic response, we examine how the pressure at the membrane crests changes with increasing $q$, as shown in~\cref{fig:results_fig}--(b) (see SM Sec.~5.2 for details on the pressure derivation).
In the case of the strict two-dimensional membrane, we find that the pressure monotonically increases as the fluctuation wavelength decreases. 
However, when the finite thickness nature of a membrane is considered, the pressure exhibits a maximum for $q<q_2$, then changes sign at $q = 2/\ell = 100$. 
This pressure reversal phenomenon indicates that flow reversal occurs in the vicinity of the membrane.
Because the flows change direction, we observe the appearance of two kinds of stagnation points, as shown in \cref{fig:results_fig}~(a)--(ii) and SM~Sec.~6.3. 
We label the two as `circulatory' and `extensional' flow stagnation points based on the flow patterns observed in \cref{fig:results_fig}~(a)--(ii).

\par

We characterize the nature of the two stagnation points by examining the strain rate tensor $D_{ij} = (\partial_i u_j + \partial_j u_i)/2$. 
For circulatory stagnation points, the local strain rate vanishes, indicating that these points do not contribute to the viscous dissipation responsible for dynamic slowing in the thickness-dominated regime (see Sec.~6.3 SM). 
However, at the other stagnation points, the strain rate tensor has non-zero diagonal components $D_{11} = -D_{22} = \dot{\epsilon}$, indicating that the flows at these points are extensional in nature.
We find that the extensional points are located at
\begin{equation}
    (x,z)^\pm_{\mathrm{ext}} = \bigg((2n+1)\frac{\pi}{2q}, \pm\frac{\ell}{2}\pm \frac{2}{q(\ell q - 2)}\bigg),\quad n\in\mathbb{Z}~,
\end{equation}
and their strain rate is given by
\begin{equation}
    \dot{\epsilon} = \frac{1}{2}\omega h_0(\ell q - 2)\exp\Big(\frac{-2}{\ell q - 2}\Big)~.
\end{equation} 
\Cref{fig:results_fig}~(c) shows $\dot{\epsilon}$ and $z^+_{\mathrm{ext}}$ for $n=0$ in terms of the wavenumber $q$.
When $q \sim q_2$, the stagnation points appear with negligible strain rate infinitely far away from the membrane surface. 
With increasing $q$, the stagnation points become localized near the membrane surfaces, and exhibit increasingly large extensional strain rates. 
In this large wavenumber regime, in-plane deformation gradients at the membrane surfaces couple to the bulk fluid and give rise to the observed stagnation point phenomena.\par
\textit{Discussion}--- 
Previously, we showed that explicitly accounting for the membrane thickness modifies the coupling between the membrane and surrounding bulk fluids, resulting in modified dynamics of bending mode relaxation (\cref{fig:intro_fig}~(b)).
As a result, fluctuations with $q > q_2 = 2/\ell$ exhibit pressure inversion at the membrane surfaces~(\cref{fig:results_fig}~(b)), which is marked by the development of vortices and local stagnation points associated with circulatory and extensional flows~(\cref{fig:results_fig}~(c)).
In particular, the extensional flow points display a mode of bulk fluid viscous dissipation caused by the compression and extension of the finite thickness membrane during bending~(\cref{fig:intro_fig}~(c)--(ii)). 
This mode contributes to a decrease in the ability of the bulk fluid to dissipate membrane fluctuations and thus leads to the dynamic slowing down of bending modes in the ``thickness'' regime.

 While we do not explicitly model lipid rotations due to our treatment of the bilayer as a continuous body, we interpret surface compression and extension in terms of the schematic depictions, i.e. \cref{fig:intro_fig}--(b), which illustrate the inferred lipid reorientation and displacements associated with the modes captured by our model.
Our focus in this work revolves around the emergent relaxation phenomena resulting from the coupling between membrane mechanics and environment hydrodynamics in the thickness regime.
Thus, we expect that if one explicitly includes the elastic contributions due to lipid tilt~\cite{hamm2000elastic,terzi_consistent_quad_2019,terzi2017novel,pinigin2020additional}, this effect will manifest itself in the conservative part of the dispersion relation, as discussed shortly.

Experimentally, membrane dynamics on nanometer length scales are often probed through density fluctuations via structure factor measurements~\cite{Zilman2002, yi_bending_2009, nagao_observation_2009, watson2011intermediate, kelley_nanoscale_2023,Faizi2024}.
Such analyses require a statistical description of membrane relaxation across a wide range of length scales, including those at the membrane thickness.
The dispersion relation $\omega(q)$ of~\cref{eq:dispersion_relation} connects the current work to the statistics of membrane fluctuations, which are measured experimentally via techniques such as neutron spin echo (NSE), dynamic light scattering, or flicker spectroscopy~\cite{monzel_measuring_2016, Pecreaux2004, brochard_erythrocytes_1975,Faizi2024}.
To link $\omega(q)$ with the dynamics of fluctuating membranes, we express it as a ratio between conservative and dissipative parts as
\begin{equation}\label{eq:disp_visc_elastic_separated}
    \omega(q) = -\frac{\kappa^\text{eff}_{\boldsymbol{q}}}{\zeta^\text{eff}_{\boldsymbol{q}}} 
    = 
    -\frac{\left(\frac{1}{2}q^4 + \Gamma q^2\right)}{\mathrm{Ca}\Gamma q\left(4+q^2 \ell^2\right)}~,
\end{equation}
where $\kappa^\text{eff}_{\boldsymbol{q}} \equiv \frac{1}{2}q^4 + \Gamma q^2$ represents the elastic response of the membrane due to out-of-plane bending and tension phenomena, and $\zeta^\text{eff}_{\boldsymbol{q}} \equiv\mathrm{Ca}\Gamma q\left(4+q^2 \ell^2\right)$ encodes all dissipative processes associated with coupling to the bulk fluid.
 If additional microscopic phenomena, such as lipid tilt, are included in the Hamiltonian description for deriving the conservative forces of \cref{eq:shape_eqn_incompressible}, they would manifest in $\kappa^\text{eff}_{\boldsymbol{q}}$.
Here, we choose to proceed with the simplest possible description of membrane energetics to understand how the thickness effects alter the hydrodynamic coupling of the membrane with its surroundings.

The form of \cref{eq:disp_visc_elastic_separated} motivates us to model the evolution of height fluctuations $h_\mathbf{q}$ through an effective overdamped Langevin equation as
\begin{equation}\label{eq:height_modes_ode}
    0 = -\zeta^\text{eff}_{\boldsymbol{q}}\partial_t \four{h}(t)  - \kappa^\text{eff}_{\boldsymbol{q}}\four{h}(t) + f_{\boldsymbol{q}}\left(t\right)~, 
\end{equation}
where $f_\mathbf{q}(t)$ is a Gaussian distributed random force with $\left<f_\mathbf{q}(t)\right> =0$ and $\left<f_\mathbf{q}(t)f_\mathbf{-q}(t')\right> = 2k_{\mathrm{B}}T \zeta^{\mathrm{eff}}_{\boldsymbol{q}}\delta\left(t-t'\right)$ \cite{Granek1997}. 
 Reinterpreting our results in the context of statistical mechanics, we can treat $\kappa^\text{eff}_{\boldsymbol{q}}$ as an effective membrane spring constant and $\zeta^\text{eff}_{\boldsymbol{q}}$ as an effective friction due to the coupling between the membrane and the surrounding fluid.
Through the lens of the fluctuation-dissipation theorem~\cite{kubo1966fluctuation,Granek1997,Lacoste2005}, the balance between these two terms dictates how thermal fluctuations relax.  
At high wavenumbers ($q > q_2$), bulk shear forces at the membrane surfaces introduce an additional dissipation mechanism that modifies $\zeta^\text{eff}_{\boldsymbol{q}}$, leading to altered relaxation dynamics.  

The Langevin interpretation in \cref{eq:height_modes_ode} suggests that measurable fluctuation spectra may deviate from predictions based on strictly two-dimensional membrane models, particularly at nanometer length scales.
Based on these insights, we expect measurements for vesicles with radius $L = 200~\mathrm{nm}$, which leads to $\mathrm{Ca} = 1$ and $\Gamma = 0.65$, to exhibit thickness effects on the relaxation dynamics for scattering with $\lambda_{\mathrm{scatter}} \lesssim 2\pi L / q_2 \approx 12~\mathrm{nm}$.
The timescale associated with this length scale is $\tau = \mu^{\mathrm{b}} \lambda_{\mathrm{scatter}}^3/\Gamma k_\mathrm{b} \approx 40~\mathrm{ns}$ for the values used in \cref{fig:intro_fig}.
 We note, however, that the relevant wavelengths in the thickness regime, i.e., $\lambda \lesssim 10$ nm, approach only a few times the bilayer thickness. 
At such short scales, molecular details that lie outside the present continuum framework (e.g., lipid tilt and local chemical heterogeneities) may become important.
Thus, our predictions should be interpreted with caution in the aforementioned regime. 
Nevertheless, for large unilamellar vesicles of radius $L \sim 200$~nm, the associated relaxation timescales can fall into the $\mathcal{O}(1$–$10$ ns) window that is accessible to NSE experiments~\cite{yi_bending_2009,nagao2023neutron}. 
Since NSE spectra reflect integrated contributions from many membrane modes rather than a single $q$, signatures of the thickness regime may still 
manifest indirectly through the parameters extracted from fits to intermediate scattering functions~\cite{Zilman1996,Zilman2002}. 
Thus, while the present model does not attempt a direct one-to-one comparison with specific experiments, it identifies a dynamical mechanism that could be probed by NSE, albeit with the above limitations in mind.
Finally, we note that different experimental techniques probe distinct length and time scales and rely on varied measurement principles~\cite{watson2011intermediate}.
Consequently, care is required when relating measurements from different experimental characterization methods, such as NSE, dynamic light scattering, and flicker spectroscopy, as each provides complementary but not identical views of membrane fluctuations.
This diversity highlights the importance of interpreting theoretical predictions in the context of the specific experimental method used.

In addition to spectroscopic methods, recent advancements in high-resolution particle tracking velocimetry~\cite{tanaka2024effect,kazoe2021super} offer promising opportunities for visualizing the effects of the high wavenumber flows (\cref{fig:results_fig}--(a)-(i)) on solute transport in the vicinity of the membrane. 
A potential experimental setup would track the motion of a particle in the vicinity of a supported or free-standing lipid bilayer. 
By studying the modes of the correlations in particle trajectories $\boldsymbol{R}(t)$, one could confirm the conditions under which the novel mode of viscous dissipation would be realized experimentally. 
On the computational side, molecular dynamics methods can also provide insight into the membrane fluid coupling observed in the thickness regime. 
A recent method~\cite{sadeghi2021hydrodynamic} combines analytic results for bulk hydrodynamics with a particle description for the membrane, where each particle describes a coarse-grained membrane patch, enabling accurate analysis of high frequency membrane processes.
By including the analytic results for the flow fields in this work (Sec.~5.3 SM), this approach provides a way to recover and validate the height mode fluctuations governed by the dispersion relation. 

 Previous modeling efforts on lipid bilayers have employed modified Helfrich approaches to incorporate thickness effects into membrane mechanics~\cite{seifert1993viscous, watson2011intermediate,fournier2015hydrodynamics, rahimi2012shape,galimzyanov2020monolayerwise,deseri2008derivation,evans1994hidden,terzi2017novel,Levine_DeterminantsBending_2014,hamm2000elastic,terzi_consistent_quad_2019}. 
 Several of these models incorporate lipid tilt---i.e., deviations of lipid orientation from the local membrane normal---as an additional elastic degree of freedom. 
Interestingly, the high-$q$ scaling obtained with our model for the dispersion relation, $\omega(q)\sim q$, is similar to that found in models that incorporate lipid tilt. 
In tilt models, the static fluctuation spectrum takes the form~\cite{terzi_consistent_quad_2019,Levine_DeterminantsBending_2014,terzi2017novel}
\begin{equation}
\left< |h_{\boldsymbol{q}}|^2 \right> = \frac{k_B T}{\frac{1}{2}k_{\mathrm{b}} q^4} + \frac{k_B T}{\frac{1}{2}k_{\mathrm{t}} q^2}
=
\frac{ 2 k_B T}{q^4} \cdot \frac{ k_{\mathrm{t}}^\mathrm{o} + q^2 }{ k_{\mathrm{t}} }
~,
\end{equation}
where $k_{\mathrm{b}}$, $k_{\mathrm{t}}$, and $k_{\mathrm{t}}^\mathrm{o} \equiv k_{\mathrm{t}}/k_{\mathrm{b}}$ are the bending and tilt moduli, along with their ratio, respectively. 
Taking the inverse of the fluctuation spectrum yields an effective membrane elastic modulus 
\begin{equation}
\kappa_q^\mathrm{eff} \propto \left< |h_{\boldsymbol{q}}|^2 \right>^{-1} \sim \frac{k_{\mathrm{t}} q^4}{ k_{\mathrm{t}}^\text{o} +  q^2 }~,
\end{equation}
where we find that $\kappa_q^\mathrm{eff} \sim k_\text{b} q^4$ for $q \ll k_t^\text{o}$ and $\kappa_q^\mathrm{eff} \sim k_{\mathrm{t}} q^2$ for $q \gg k_t^\text{o}$, reproducing the crossover from the bending- to the tilt-dominated regime. 
If we model dissipation via a Stokes drag kernel that scales as $\sim 1/q$, we find $\omega(q) \sim q$ for $q \gg k_t^\text{o}$, which is similar to our predictions of the dispersion relation, i.e., \cref{eq:disp_visc_elastic_separated}.
While this provides a simple qualitative argument about a naive inclusion of tilt effects, we note that for $q \sim 2\pi/\delta$, the hydrodynamic response encoded in the denominator of \cref{eq:dispersion_relation,eq:disp_visc_elastic_separated} will likely be altered by finite thickness effects due to the membrane–fluid coupling occurring at the bounding surfaces.
This is the case with our model where even in the asymptotic, i.e. $\delta \ll L$, and linear response limits we consider, the elastic response remains unaltered from the two-dimensional result, and finite thickness effects appear via the hydrodynamic response.

Therefore, in our model of \cref{eq:disp_visc_elastic_separated}, the linear in $q$ scaling results from a purely dissipative mechanism.
Particularly, the elastic response is governed by pure bending and tension, while $\zeta_q^\mathrm{eff}$ includes a term from surface shear as a consequence of the membrane's finite thickness.
This contrast emphasizes that similar dispersion relations can emerge from physically distinct origins--one elastic, the other hydrodynamic--and that interpreting relaxation dynamics requires careful consideration of all relevant physics at play.
The present model assumes lipids remain rigid and aligned locally with the mid-surface. 
Including tilt would require additional higher-order corrections beyond the current formulation.
In order to do so, one would need to begin with the $(2+\delta)$-dimensional formalism~\cite{omar2023electrostatics,omar2023balance,omar2025constitutive} and derive equations for the evolution of a director field of higher order than the normal. 
In the linear response limit we study here, this likely would only renormalize the elastic contributions as the viscous coupling between the membrane and fluid would remain unaltered.

 Traditionally, the most common approach to incorporating finite thickness effects in membrane dynamics is the intermonolayer slip (IS) model~\cite{evans1994hidden,yeung1995unexpected,seifert1993viscous}, which describes the membrane as two coupled monolayers that can slide past one another. Dissipation arises from relative motion between these leaflets, governed by a phenomenological friction coefficient. Both the IS model and the present $(2+\delta)$-dimensional theory predict in-plane flows at the membrane surfaces~\cite{fournier2015hydrodynamics}, but their origin is physically different. 
In the IS model, surface flows emerge from monolayer area asymmetry and density relaxation (see SM Fig.9(a)–(c)), whereas in our framework, they arise from lipid reorientation and interfacial shear generated by three-dimensional hydrodynamic coupling (Fig.\ref{fig:intro_fig}--(b)).

The differences in dissipation mechanisms lead to distinct dispersion scalings. The IS model predicts $\omega(q) \sim q^2$ at intermediate wavenumbers, where monolayer asymmetries relax slowly~\cite{watson2011intermediate,rodriguez2009bimodal,sadeghi2021hydrodynamic}. However, at higher wavenumbers, this scaling reverts to $\omega(q) \sim q^4$, indistinguishable from the classical Helfrich limit. As such, IS becomes effectively invisible in the nanoscopic regime, where its contribution simply renormalizes the bending modulus. In contrast, the $(2+\delta)$-dimensional theory captures a qualitatively distinct dissipation mechanism at short wavelengths. For $q > q_2 = 2/\ell$, interfacial shear dominates energy loss, yielding a high-$q$ dispersion of $\omega(q) \sim q$ (see SM Fig.~9(d)). This behavior cannot be captured by IS and represents a fundamentally different mode of membrane–fluid coupling.

A further distinction lies in how each model handles geometry. The IS model, despite referencing bilayer thickness, treats the membrane as physically two-dimensional, imposing boundary conditions at $z = 0^{\pm}$. This approximation holds at long wavelengths but becomes problematic when the relevant length scales approach the bilayer thickness, where accurate evaluation of surface stresses is critical. The present theory resolves this limitation by enforcing boundary conditions at the actual membrane–fluid interfaces, enabling a consistent treatment of interfacial shear and resultant flows near the membrane.

While distinct in formulation and regime of validity, the IS and $(2+\delta)$-dimensional models are not mutually exclusive. Rather, they describe complementary modes of dissipation: IS captures monolayer slip at intermediate $q$, and our model captures interfacial shear at high $q$. 
Specifically, we can estimate the crossover wavenumbers associated with IS dissipation using representative parameters from SM Fig.~9 in Eq.~(7) in Ref.~\cite{seifert1993viscous}. These define a regime bounded between $q_1^\mathrm{IS} L \approx 3.2$ and $q_2^\mathrm{IS} L\approx 28$.  Thus, the dissipative mechanism discussed in the present work is relevant in a complementary wavenumber regime $q \geq q_2$, where $q_2>q_2^{IS}$ (see \textit{Results}). 
Therefore, a promising direction would be to combine these approaches, producing a unified theory of bilayer dynamics that remains accurate across all relevant length scales.

Regarding the relevance of our findings on biological function, we expect that they have implications for processes such as mechanosensitive protein function and intercellular contact.
For instance, it has recently been observed that increased membrane thickness correlates with reduced permeability to solute molecules~\cite{frallicciardi_membrane_2022}. 
While multiple factors likely contribute to this effect, our results suggest that nanoscale circulations induced by finite thickness effects could potentially influence local mixing near the membrane. 
Such flows may, in turn, affect solute concentration gradients, which could play a role in permeability regulation. 

The effects on local solute concentration gradients would be particularly important during neuronal signaling, where ion concentrations at the membrane interface play a central role in action potential propagation~\cite{hodgkin1949effect,hille1978ionic,tyagi2025sculpting}. 
While prior work has modeled ion diffusion and permeability~\cite{row2025spatiotemporal} for finite thickness membranes, incorporating the hydrodynamic coupling of the present work would yield a more complete picture of how bilayer properties influence neural function.
Additionally, the predicted pressure inversion phenomenon at nanoscales,~\cref{fig:results_fig}-(b), can lead to hydrodynamic-related forces~\cite{Bartolo2002} that would mediate intercellular contact and affect intermembrane junction fluctuations~\cite{liu2019hydrodynamics, lin_analysis_2006} and organelle interactions.
 For instance, during iron-driven cell death~\cite{dos2025organelle}, the endoplasmic reticulum and mitochondria interact at nanometer scales. Finite thickness hydrodynamics could facilitate these contacts by inducing attractive or circulating forces. Future studies incorporating confinement and adhesion may reveal the extent to which these effects contribute to intermembrane dynamics.
Further experimental studies on free-standing and confined membranes will also help reveal the relevance of finite thickness effects on the discussed processes.

\textit{Conclusion}---In summary, we have shown that the finite thickness of lipid bilayers affects how they couple with the surrounding media. In particular, shear forces caused by lipid reorientations during bending develop at the membrane surfaces, resulting in in-plane flows that are not observed in strictly two-dimensional membrane models. This mechanism leads to a redistribution of momentum between the membrane and bulk fluid, where in-plane lipid motion generates shear stresses that induce circulatory and extensional flow patterns in the surrounding fluid. For fluctuations on the order of the membrane thickness, we show that this coupling results in pressure inversion at the membrane surfaces, indicating the existence of flow reversal in the vicinity of the bilayer. Stagnation points associated with bulk fluid circulatory and extensional flows mark where the flow reversal occurs.
Finally, we have shown that our dispersion relation provides a direct link between continuum membrane mechanics and measurable fluctuation spectra, offers predictions testable by neutron spin echo or high-resolution velocimetry, and complements existing membrane theories through comparison across physical regimes.

\paragraph{Acknowledgments}
{\small ZGL and YADO would like to acknowledge discussions with Kranthi K. Mandadapu.
ZGL and YADO were financially supported by the University of California, Berkeley and the Director, Office of Science, Office of Basic Energy Sciences, of the U.S. Department of Energy under contract No. DEAc02-05CH11231.
ZGL was supported partially by the National Science Foundation through NSF-DFG 2223407 and the Deutsche Forschungsgemeinschaft (German Research Foundation)—509322222. 
DF (dfrag) acknowledges Princeton University and the Department of Chemical and Biological Engineering for financial support provided through their start-up funding.}

\nocite{Bassereau2014,Aris1989,Tsafrir2003,gurtin2010mechanics,boyd2001chebyshev,Purcell1977,deen_transport,Lidmar2003,Doi1988,Doi2011}

\bibliography{Lin_Resp_Fin_Thick_Mems/bibliography/ref_corrected}
\end{document}